\documentclass[12pt]{article}
\pdfoutput=1
\usepackage[letterpaper,margin=1in,headsep=0.1in,headheight=1in]{geometry}

\usepackage[T1]{fontenc}
\usepackage{ae,aecompl}
\usepackage[utf8]{inputenc}
\usepackage{charter} 

\usepackage[hidelinks]{hyperref}
\usepackage{graphicx}
\usepackage[sort&compress,numbers]{natbib}

\usepackage{amsmath,amssymb,amstext}
\usepackage{xspace}
\usepackage{aas_macros}

\hypersetup{pdfauthor={Mao, Peter et al.},pdftitle={Vera C. Rubin Observatory as a Flagship Dark Matter Experiment}}

\newcommand\snowmass{
\begin{center}
  \rule[-0.2in]{\hsize}{0.01in}\\
  \rule{\hsize}{0.01in}\\
  \vskip 0.1in
  Submitted to the Proceedings of the US Community Study\\ 
  on the Future of Particle Physics (Snowmass 2021)\\
  \rule{\hsize}{0.01in}\\
  \rule[+0.2in]{\hsize}{0.01in}\\[-2em]
\end{center}
}

\usepackage[firstpage=true]{background}
\backgroundsetup{contents={\parbox{6.5in}{\snowmass}}, scale=1,placement=top,opacity=1,color=black,position={3.25in,1.2in}}

\usepackage{fancyhdr}
\fancypagestyle{plain}{%
  \fancyhf{}%
  \fancyhead[C]{}
  \fancyfoot[C]{\thepage}
}

\fancypagestyle{empty}{%
  \fancyhf{}%
  \fancyhead[C]{\textit{Vera C.\@ Rubin Observatory as a Flagship Dark Matter Experiment}}
  \fancyfoot[C]{\thepage}
}

\usepackage{titling}
\title{Snowmass2021 Cosmic Frontier White Paper:\\{}
Vera C.\@ Rubin Observatory as a Flagship Dark Matter Experiment}
\date{}

\usepackage{authblk}

\author[1]{Yao-Yuan~Mao}
\author[2]{Annika~H.~G.~Peter}
\author[3,4]{Susmita~Adhikari}
\author[5]{Keith~Bechtol}
\author[6]{Simeon~Bird}
\author[7]{Simon~Birrer}
\author[8]{Jonathan~Blazek}
\author[9]{Jeffrey~L.~Carlin}
\author[10]{Nushkia~Chamba}
\author[11,12]{Johann~Cohen-Tanugi}
\author[13]{Francis-Yan~Cyr-Racine}
\author[14,15]{Tansu~Daylan}
\author[13]{Birendra~Dhanasingham}
\author[3,16,17]{Alex~Drlica-Wagner}
\author[18]{Cora~Dvorkin}
\author[19]{Christopher~Fassnacht}
\author[1]{Eric~Gawiser}
\author[20]{Maurizio~Giannotti}
\author[21]{Vera~Gluscevic}
\author[22,23]{Alma~Gonzalez-Morales}
\author[24]{Ren\'ee~Hlo\v{z}ek}
\author[19,25]{M.~James~Jee}
\author[26]{Stacy~Kim}
\author[27]{Akhtar~Mahmood}
\author[28]{Rachel~Mandelbaum}
\author[18,29,30]{Siddharth~Mishra-Sharma}
\author[31]{Marc~Moniez}
\author[21,32]{Ethan~O.~Nadler}
\author[33]{Chanda~Prescod-Weinstein}
\author[19]{J.~Anthony~Tyson}
\author[7,34]{Risa~H.~Wechsler}
\author[6]{Hai-Bo~Yu}
\author[35]{Gabrijela~Zaharijas}
\affil[1]{Department of Physics and Astronomy, Rutgers, The State University of New Jersey, Piscataway, NJ 08854, USA}
\affil[2]{CCAPP, Department of Physics, Department of Astronomy, The Ohio State University, Columbus, OH 43210, USA}
\affil[3]{Department of Astronomy and Astrophysics, University of Chicago, Chicago, IL 60637, USA}
\affil[4]{Department of Physics, Indian Institute of Science Education and Research, Pashan, Pune 411008, India}
\affil[5]{Department of Physics, University of Wisconsin--Madison, 1150 University Avenue, Madison, WI 53706, USA}
\affil[6]{Department of Physics and Astronomy, University of California, Riverside, CA 92521, USA}
\affil[7]{Kavli Institute for Particle Astrophysics and Cosmology and Department of Physics, Stanford University, Stanford, CA 94305, USA}
\affil[8]{Department of Physics, Northeastern University, Boston, MA, 02115, USA}
\affil[9]{AURA/Vera C. Rubin Observatory, 950 N Cherry Ave, Tucson, AZ 85719 USA}
\affil[10]{The Oskar Klein Centre, Department of Astronomy, Stockholm University, AlbaNova, SE-10691 Stockholm, Sweden}
\affil[11]{LUPM, University of Montpellier and CNRS, France}
\affil[12]{LPC, Universit\'e Clermont-Auvergne and CNRS, France}
\affil[13]{Department of Physics and Astronomy, University of New Mexico, Albuquerque, NM 87106, USA}
\affil[14]{Department of Physics and Kavli Institute for Astrophysics and Space Research, Massachusetts Institute of Technology, Cambridge, MA 02139, USA}
\affil[15]{Department of Astrophysical Sciences, Princeton University, Peyton Hall, Princeton, NJ 08544, USA}
\affil[16]{Fermi National Accelerator Laboratory, P. O. Box 500, Batavia, IL 60510, USA}
\affil[17]{Kavli Institute for Cosmological Physics, University of Chicago, Chicago, IL 60637, USA}
\affil[18]{Department of Physics, Harvard University, Cambridge, MA 02138, USA}
\affil[19]{Department of Physics and Astronomy, University of California, Davis, CA  95616, USA}
\affil[20]{Department of Chemistry and Physics, Barry University, 11300 NE 2nd Ave., Miami Shores, Florida 33161, USA}
\affil[21]{Department of Physics \& Astronomy, University of Southern California, Los Angeles, CA, 90007, USA}
\affil[22]{Consejo Nacional de Ciencia y Tecnolog\'ia, Av. Insurgentes Sur 1582. Colonia Cr\'edito Constructor, Del. Benito Jurez C.P. 03940, M\'exico D.F. M\'exico}
\affil[23]{Departamento de F\'isica, Divisi\'on de Ciencias e Ingenier\'ias, Campus Le\'on, Universidad de Guanajuato, Le\'on  C.P. 37150. Guanajuato, M\'exico}
\affil[24]{Dunlap Institute for Astronomy and Astrophysics \& David A. Dunlap Department of Astronony and Astrophysics, University of Toronto, 50 St George Street, Toronto ON M5S3H4, Canada}
\affil[25]{Department of Astronomy, Yonsei University, Seoul, 03722, South Korea}
\affil[26]{Department of Physics, University of Surrey, Guildford, GU2 7XH, United Kingdom}
\affil[27]{Department of Physics, Bellarmine University, Louisville, KY 40205, USA}
\affil[28]{McWilliams Center for Cosmology, Department of Physics, Carnegie Mellon University, Pittsburgh, PA 15213, USA}
\affil[29]{The NSF AI Institute for Artificial Intelligence and Fundamental Interactions, Cambridge, MA 02139, USA}
\affil[30]{Center for Theoretical Physics, Massachusetts Institute of Technology, Cambridge, MA 02139, USA}
\affil[31]{Laboratoire de physique des 2 infinis Ir\`ene Joliot-Curie, CNRS Universit\'e Paris-Saclay, B\^at. 100, Facult\'e des sciences, F-91405 Orsay Cedex, France}
\affil[32]{Carnegie Observatories, 813 Santa Barbara Street, Pasadena, CA 91101, USA}
\affil[33]{Department of Physics and Astronomy, University of New Hampshire, Durham, NH 03824 USA}
\affil[34]{SLAC National Accelerator Laboratory, Menlo Park, CA 94025, USA}
\affil[35]{Center for Astrophysics and Cosmology, University of Nova Gorica, Slovenia}

\begin{document}

\maketitle

\begin{abstract}

Establishing that Vera C. Rubin Observatory is a flagship dark matter experiment is an essential pathway toward understanding the physical nature of dark matter. 
In the past two decades, wide-field astronomical surveys and terrestrial laboratories have jointly created a phase transition in the ecosystem of dark matter models and probes. 
Going forward, any robust understanding of dark matter requires astronomical observations, which still provide the only empirical evidence for dark matter to date.
We have a unique opportunity right now to create a dark matter experiment with Rubin Observatory Legacy Survey of Space and Time (LSST). 
This experiment will be a coordinated effort to perform dark matter research, and provide a large collaborative team of scientists with the necessary organizational and funding supports. This approach  leverages existing investments in Rubin. 
Studies of dark matter with Rubin LSST will also guide the design of, and confirm the results from, other dark matter experiments. 
Supporting a collaborative team to carry out a dark matter experiment with Rubin LSST is the key to achieving the dark matter science goals that have already been identified as high priority by the high-energy physics and astronomy communities. 

\end{abstract}


\pagestyle{empty}

\section{Wide-field Surveys as Dark Matter Experiments}

More than 85 years after its astrophysical discovery, the fundamental nature of dark matter remains one of the foremost open questions in physics.
Over the last several decades, an extensive experimental program has sought to determine the cosmological origin, fundamental constituents, and interaction mechanisms of dark matter. 
While this experimental program has historically focused on weakly-interacting massive particles (WIMPs), there is strong, growing theoretical motivation to explore a broader set of dark matter candidates.
As the high-energy physics program expands to ``search for dark matter along every feasible avenue'' \citep{P5Report}, it is essential to remember that the only direct, empirical measurements of dark matter properties to date come from astrophysical and cosmological observations.

Two decades ago, the cold dark matter (CDM) phenomenological model (QCD axions and WIMPs are almost synonymous with CDM) reigned supreme in the astronomy community.  However, there were hints that dark matter may have more interesting particle and cosmological phenomenology \cite{moore1999a,klypin1999,deBlok:2001hbg}.  On the particle side, most effort was focused on searches for WIMPs and QCD axions in terrestrial laboratories (see, e.g., \cite{2005PhR...405..279B}).  Since then---and especially since the last Snowmass process in 2013---the landscape of particle dark matter models and experimental search techniques has grown tremendously.  This growth is apparent in the \emph{US Cosmic Visions: New Ideas in Dark Matter: 2017 Community Report} white paper \cite{Battaglieri2017} and the numerous white papers submitted as part of the current Snowmass process.

What happened to change this landscape?  First, there has not yet been a discovery of supersymmetry or QCD axions at the LHC or via direct and indirect detection.  Second, the advent of deep, wide-field optical surveys two decades ago has enabled new opportunities to test dark matter physics.  Although the largest contemporary optical surveys are optimized for dark energy science, they are showing themselves to be drivers for dark matter science, too.  Astronomical anomalies uncovered in a wide range of data sets have encouraged a broader range of theoretical modeling and deeper studies of astrophysical phenomenology. The broad range of dark matter models that can be tested with astrophysical/cosmological observations is more widely understood than twenty years ago.

There are now many precision tests of dark matter physics with astronomical observations.  Weak gravitational lensing links galaxies to halos, enabling the community to use moderate-sized galaxies and above to measure the halo mass function (e.g., \cite{wittman2000,Mandelbaum:2006}).  The discovery of a new class of tiny galaxy--the ultra-faint dwarf galaxies--sharpened tests of dark matter on subgalactic scales (e.g., \cite{Simon190105465,Nadler200800022,Kim:2021zzw}).  The study of high-redshift quasars has facilitated tests of the halo mass function across much of the age of the universe via measurements of the Lyman alpha forest (e.g., \cite{2016JCAP...08..012B}).  The discovery of dozens of strongly lensed galaxies and quasars has permitted tests of dark matter on even smaller scales \cite{Oguri:2010,Nierenberg:2019pdj}.  Recently, newly discovered stellar streams of the Milky Way show the promise of using the number, orbits, and density function of streams to test dark matter physics on scales below the mass threshold at which galaxies form, as with substructure lensing \cite{DES:2018imd,Banik:2019cza}.  Interestingly, even smaller halos may be discovered via strong-lensing caustics \cite{Dai:2020rio}.  The diversity, sensitivity, and precision of these probes allow us to robustly test a wide variety of dark-matter candidates, and any ``anomaly" discovered with a telescope inspires new particle candidates and tests thereof.\footnote{A more comprehensive description of cosmic probes of dark matter can be found in the set of white papers solicited by the Snowmass topical group on ``Dark Matter: Cosmic Probes'' \cite{DMhalosWP,DMextremeWP,DMPBHWP,DMfacilitiesWP,DMDESIWP,DMCMB64WP,TFastroprobesWP}.}
Wide-field surveys should thus be thought of as \emph{dark matter experiments}.

Wide-field survey dark-matter experiments support and enhance other dark-matter experiments and searches.  Consider the case of the conventional WIMP.  The discovery of ultra-faint dwarf galaxies around the Milky Way and the characterization of their density profiles with follow-up spectroscopy enabled precision tests of the WIMP annihilation cross section with the Fermi-LAT experiment.  These measurements now exclude WIMP masses below 100 GeV for several annihilation channels \citep{1503.02641}, while WIMPs with other particle properties can be tested with direct-detection and collider experiments.  Furthermore, the abundance of ultra-faint dwarf galaxies and their relatively high central densities are in line with precision predictions for WIMP/CDM via high-resolution $N$-body simulations, excluding wide swathes of non-WIMP (e.g., sterile neutrino, ultra-light dark matter) parameter space \cite{errani2018,Nadler200800022,Kim:2021zzw}.  

Laboratory and telescope-based probes continue to collaborate to provide the most robust constraints on particle dark matter candidates, most notably sterile neutrino and hidden-sector candidates. Any future detection of dark matter from laboratories will also need to be confirmed by telescope-based observations to ensure the detection is indeed the astrophysical dark matter. Hence, creating a high-quality map between the microphysical properties of dark matter and dark matter's astrophysical signals (e.g., cosmological clustering) is a key priority for this program.

Importantly, if the dark sector has little interaction with the Standard Model, then observational cosmology may be our only hope of measuring the particle properties of this sector \cite{BuckleyPeter:2017}.  Several of the dark-matter candidates identified in the 2014 P5 report  (e.g., sterile neutrinos, hidden-sector dark matter) may fall into this category.  Any interaction between dark matter and Standard Model particles will additionally imprint itself on the cosmic distribution of dark matter.  \emph{Now} is the time for a next-generation dark matter experiment using a wide-field galaxy survey.

\begin{figure}
    \centering
    \includegraphics[width=.95\textwidth]{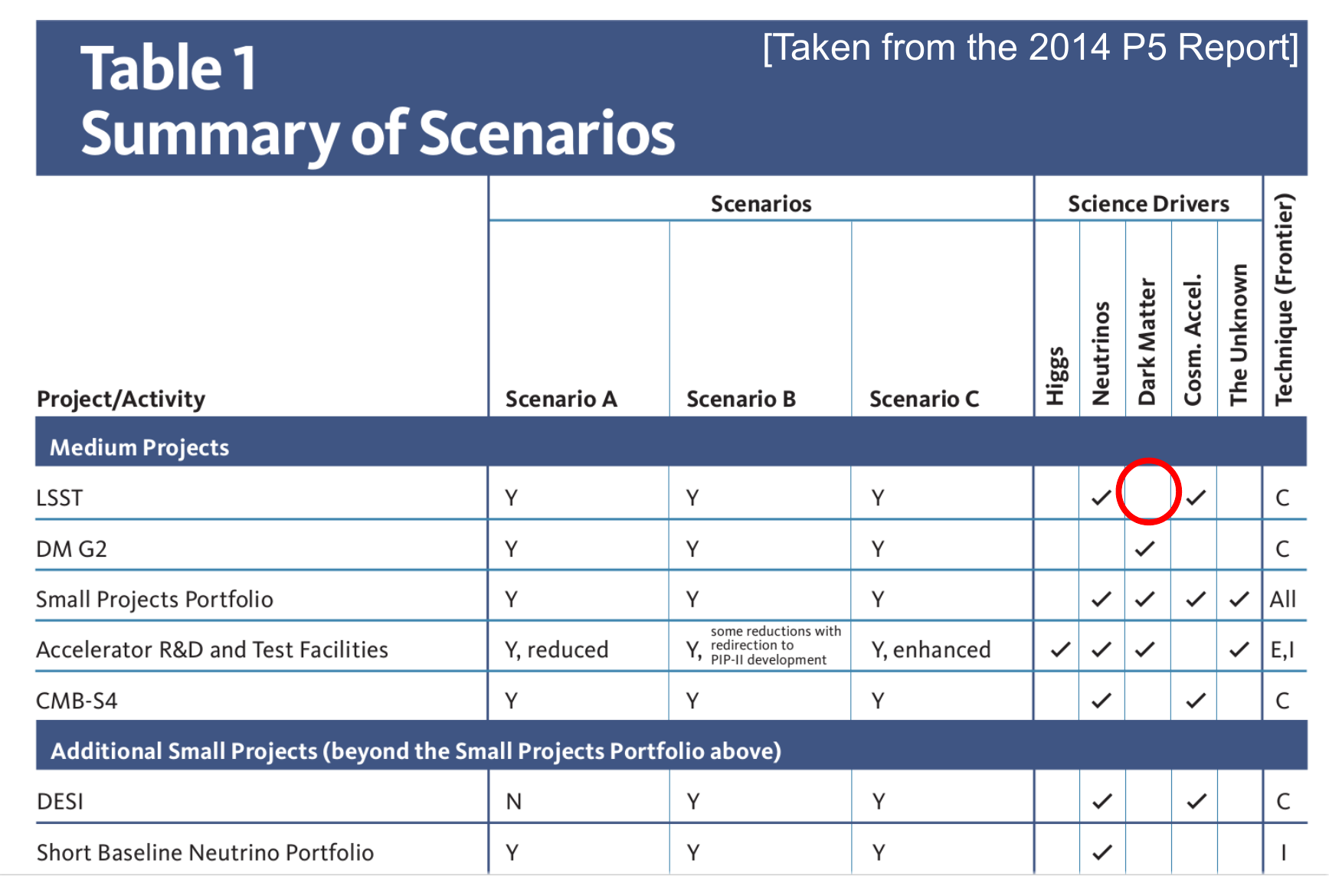}
    \caption{A reproduction of Table 1 of the 2014 Report of the Particle Physics Project Prioritization Panel, ``Building for Discovery: Strategic Plan for U.S. Particle Physics in the Global Context.'' The red circle highlights the lack of recognition that dark matter is a strong science driver for Rubin LSST, among other experiments.}
    \label{fig:p5-report}
\end{figure}

\section{A Flagship Dark Matter Experiment with the Vera C.\ Rubin Observatory}

The Vera C.\ Rubin Observatory Legacy Survey of Space and Time (LSST) will be a revolutionary wide-field survey that provides a unique and impressive platform to study dark sector physics.
In fact, the Rubin Observatory was originally envisioned as the ``Dark Matter Telescope'' \citep{Tyson:2001}.
Today, probing dark energy and dark matter is prominently listed as one of the four science pillars of the Rubin Observatory.\footnote{See, for example, \href{https://www.lsst.org/about}{www.lsst.org/about}}
However, this potential to reveal dark matter physics was not prioritized in the last P5 report (see Table~1 of 2014 P5 Report; reproduced in \autoref{fig:p5-report}). 

Dark matter science is a significant driver for the Rubin Observatory LSST, as demonstrated in Ref. \cite{drlica-wagner_2019_lsst_dark_matter}, as it has been in other wide-field surveys.
In fact, astrophysical probes provide the only constraints on the minimum and maximum mass scale of dark matter, and astrophysical observations will likely continue to guide the experimental particle physics program for years to come.

In this white paper, we further argue that establishing the Vera C.\ Rubin Observatory as a flagship dark matter experiment will bring the dark matter community together to make critical discoveries in the coming decade. Recognizing Rubin as a dark matter experiment, in a way that resembles other HEP experiments, means to provide organizational and funding support to a collaborative team working on delivering constraints on dark matter properties with Rubin LSST.
This investment is cost-effective, and will generate rich scientific and technical deliverables that benefit and complement both other dark matter experiments and other studies of fundamental physics (e.g., dark energy and neutrinos) with LSST.

\subsection{Timely Opportunities}

In the precision era of LSST, studies of dark matter and dark energy are complementary from both a technical and scientific standpoint. Understanding $\Lambda$CDM means understanding CDM and $\Lambda$ in the context of each other. Furthermore, theory and simulations have consistently shown that the macroscopic distribution of dark matter yields critical information about the microscopic physics governing dark matter. 

The study of dark matter with LSST will explore parameter space beyond the high-energy physics program's current sensitivity while being highly complementary to other experimental searches. This has been recognized in Astro2010 \citep{astro2010}, during the 2013 Snowmass Cosmic Frontier planning process \citep[e.g.,][]{1310.8642, 1310.5662, 1305.1605}, in the 2014 P5 Report \citep[]{P5Report}, and in a series of more recent Cosmic Visions reports \citep[e.g.,][]{1604.07626,1802.07216}, including the ``New Ideas in Dark Matter 2017:\ Community Report'' \citep{Battaglieri2017}.

Most recently, the Astro2020 report \citep{astro2020}, ``Pathways to Discovery in Astronomy and Astrophysics for the 2020s,'' highlights the importance of Rubin LSST in understanding the nature of dark matter. 
Sec.~1.1.2 of the report specifically mentions that 
``[t]he unknown physical natures of dark matter and dark energy, both discovered through astronomical measurements, remain outstanding grand challenges in both physics and astronomy, and great observational progress will be made in the coming decade. Addressing these profound mysteries
were prime motivations for [...] the NSF/DOE Vera C. Rubin Observatory, a wide-field 8.4\,m telescope devoted to a decade-long mapping of the entire southern sky [...].'' 
In addition, the Astro2020 Report of the Panel on Cosmology \citep[App. C,][]{astro2020} also recognizes that ``[t]he search for dark matter signatures is wide-ranging and exploratory, but the next generation of
radio telescopes for pulsar timing, \textit{large-aperture optical telescopes}, high-resolution CMB polarization mapping, GeV telescopes, and TeV-scale Cherenkov telescopes are particularly important to make
progress in this field'' (emphasis added).

\subsubsection{How will LSST Constrain Dark Matter Properties?}
\label{sec:lsst-topics}

The 2019 white paper \citep{drlica-wagner_2019_lsst_dark_matter} has detailed how Rubin LSST, an unprecedentedly powerful multi-faceted survey, has the potential to identify the fundamental constituents of dark matter through diverse observational measurements. 
A series of white papers solicited by the Snowmass topical group on ``Dark Matter: Cosmic Probes'' also highlight dark matter science with Rubin LSST \cite{DMhalosWP,DMextremeWP,DMPBHWP,DMfacilitiesWP,DMDESIWP,DMCMB64WP,TFastroprobesWP}.
We summarize the main aspects as follows.

LSST will enable studies of ultra-faint dwarf galaxies (including Milky Way satellites), stellar streams, and strong lens systems to detect and characterize the smallest dark matter halos, thereby probing the minimum mass of ultra-light dark matter, the free-streaming length of dark matter, and interactions between dark matter and Standard-Model particles. It is worth noting that dark matter--baryon interactions are currently best constrained by the Milky Way satellite abundances, and the search for Milky Way satellites will only be improved by Rubin LSST in the next 10 years.

LSST will allow precise measurements of the density profiles and shapes of dark matter halos in dwarf galaxies and galaxy clusters. The dark matter density profiles and halo shapes will be sensitive to dark matter self interactions, hence probing hidden-sector models.

LSST can surpass current limitations of microlensing measurements by directly detecting events based purely on the parallactic component of the lensing signal. The excellent photometric precision of LSST will also allow us to detect microlensing events with large impact parameters These microlensing measurements will directly probe primordial black holes and the compact object fraction of dark matter at the sub-percent level over a wide range of masses, measuring the inflationary power spectrum as well as dark matter physics.

LSST will increase the white dwarf (WD) stars census and the identification of core-collapse SNe and of their progenitors, and allow the construction of extremely accurate WD luminosity functions. Precise measurements of these stellar populations will be sensitive to anomalous energy-loss mechanisms and will constrain the properties of axions, axion-like particles, and other light and feebly interacting particles (e.g., dark photons, non-standard neutrinos).

Finally, studying dark matter with LSST allows us to explore complementarity between cosmological, direct detection, and other indirect searches for dark matter. For example, measurements of large-scale structure will spatially resolve the influence of both dark matter and dark energy, enabling searches for correlations between the two known components of the dark sector. Joint analyses with direct and indirect searches will help constrain dark matter-baryon interactions, dark matter self-annihilation, and dark matter decay. 
The different cosmic epochs probed by Rubin LSST and particle dark matter experiments allow us to distinguish primordial dark matter physics (e.g., as thermal production) from late time ones (e.g., self-interaction).

\subsection{The Need for a Collaborative Team}

Individual scientists can use LSST data to study dark matter since LSST data will be accessible to all US and Chilean scientists and later become broadly publicly available as well. However, it has been demonstrated in many cases, both in the context of large astronomical surveys and particle experiments, that a collaborative team can be the most effective to extract science from large surveys and experiments. 
In particular, dark matter science represents a unique synthesis of observational astrophysics, computational cosmology, and high-energy theory. The requirement of people with different domain knowledge working together highlights the need of a collaborative team.

More importantly, in the context of large astronomical surveys, a comprehensive understanding of complex instruments, data sets, and analysis pipelines is the foundation for any science analysis. This is the reason that modern dark energy science with wide-field surveys is almost exclusively done in large collaborations.
The same holds for dark matter science. 
Furthermore, there exist great international efforts from countries beyond the US and Chile, and some international teams have even proposed in-kind contributions to Rubin that are relevant to dark matter science. Hence, it is important that a collaborative team exist to coordinate and focus these efforts. 

To this end, the community has been self-organizing to bring together astronomers, cosmologists, and particle theorists to prepare for dark matter science using LSST through coordinated efforts in all of these areas. In early 2018, a group of experimentalists, observers, simulators, and particle theorists self-assembled to build the dark matter science case for LSST. 
With several workshops and meetings, this group published a 95-page white paper \citep{drlica-wagner_2019_lsst_dark_matter} and submitted an Astro2020 white paper \citep{astro2020_white_paper}. 
In 2019, motivated by the significant overlap in scientific interests, technical needs, and personnel between dark energy and dark matter sciences, the idea of establishing a Dark Matter Working Group as part of the Rubin LSST Dark Energy Science Collaboration (DESC) was proposed. 
This DESC Dark Matter Working Group was designed to serve as a centralized place to host this community, and to effectively leverage synergies with DESC work on the LSST data and dark energy probes, while also interacting with other relevant LSST science collaborations (e.g., Stars, Milky Way, and Local Volume; and Strong Lensing). 

\begin{figure}
    \centering
    \includegraphics[width=\textwidth]{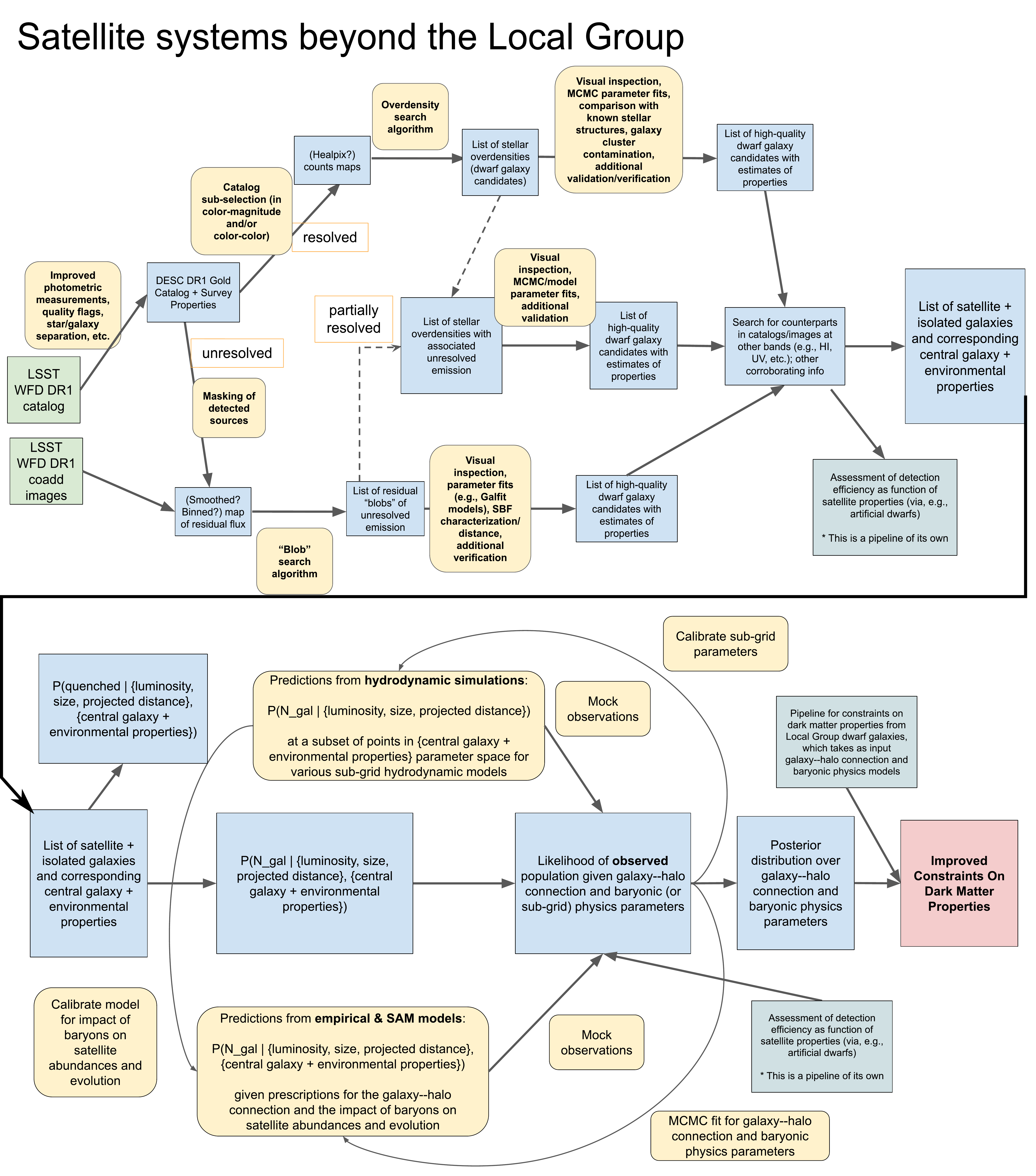}
    \caption{An example ``dark matter flowchart'' showing all the needed elements to constrain dark matter properties from LSST data for the measurement of satellite systems beyond the Local Group. This complex flowchart highlights the need for a collaborative team and organizational and funding support.}
    \label{fig:flowchart}
\end{figure}

This diverse group of scientists has already started to map out the scope of dark matter analyses: how we derive constraints from LSST data. 
The DESC Dark Matter working group has recently created a series of ``dark matter flowcharts'' that lay out the key observational and theoretical needs, and specific analysis pipelines that need to be built. 
Several flowcharts have been made to cover different analyses mentioned in Section~\ref{sec:lsst-topics}, including Milky Way satellites, satellite systems beyond the Local Group, stellar streams, strong lenses, substructures in clusters. 
Figure~\ref{fig:flowchart} shows an example of the flowchart for the measurement of satellite systems beyond the Local Group. The complex flowchart highlights the need for a collaborative team with a diverse set of expertise. 

In addition to being an example of how a community like the DESC Dark Matter Working Group can significantly improve scientific and technical productivity, this exercise has shown that the work to extract dark matter properties is large and complex. We are urgently in need of organizational and funding support to accomplish our goals.

\subsection{A Low-cost Low-risk High-reward Investment}

Recognizing Rubin LSST as a flagship dark matter experiment such that a team of scientists, such as the DESC Dark Matter Working Group, can receive organizational and funding support, similar to how direct detection, collider, and dark energy experiments operate, is the key to achieve the dark matter science goals that have already been identified as high priority by the HEP community. 

Stable funding lines from DOE have long played a critical role in community planning for and execution of sensitive dark energy measurements with dedicated facilities.  Similar to the case of pursuing dark energy science, the study of dark matter science is strengthened by a unified framework for the various types of measurements that touch on dark matter properties. These analyses will need to be done in close collaboration with the detailed work at the pixel level of the survey.  Dedicated funding for a flagship dark matter experiment with Rubin Observatory will allow us to develop better analysis frameworks together, in close connection to instrument-level work.

Most importantly, funding dark matter science with Rubin LSST is not at the expense of the success of other sciences. The cost to establish a flagship dark matter experiment with Rubin LSST is minimal, given the existing investment in dark energy science and the significant overlap of the required infrastructures and analysis methods for dark energy and dark matter. The benefit of co-locating research efforts in dark matter and dark energy has already been realized in DESC Dark Matter Working Group. By recognizing that Rubin LSST can deliver dark matter science, the DOE can ensure that the existing investment toward studies of systematics, complex analysis pipelines, and technical work at the pixel and instrument level will also encompass important dark matter science, and hence maximize the output of the existing investment. 

In addition, funding dark matter science with Rubin LSST will complement and enhance the outcome of direct detection, indirect detection, and collider experiments. LSST will provide critical information on the fundamental dark matter physics, which can either help inform the searches in particle experiments, or to collaborate with any upcoming search results from particle experiments to build a full physical picture of dark matter particles. Hence, the return of this investment is almost certainly high.

\section{Summary}

Rubin LSST will enable transformative studies of dark matter physics, as evidenced by the precursor wide-field surveys. 
In addition to advancing existing astrophysical probes of dark matter, Rubin LSST will also enable new probes of dark matter physics that have yet to be considered. 
As the particle physics community diversifies the experimental effort to search for dark matter, a critical consideration is that astrophysical observations already provide robust, empirical measurement of fundamental dark matter properties. 
Some of the astrophysical probes (e.g., merging galaxy clusters) are effectively dark matter particle collider experiments lightyears away.
In the coming decade, astrophysical observations will probe unique regions of dark matter parameter space, guide other particle experimental efforts, and connect any discoveries in a terrestrial laboratory to \emph{the} dark matter we have observed in the Universe.

Using a large astrophysical survey like Rubin LSST to study dark matter will involve a wide range of expertise, ranging from theoretical model and analysis development to detailed knowledge of the instruments. Science collaborations, such as the LSST Dark Energy Science Collaboration, enable efficient communications, and provide organizational support for a large number of scientists working together. The benefit of such collaborations has been seen in multiple DOE experiments, including direct detection, indirect detection, collider, and dark energy science. The Dark Matter Working Group within LSST DESC, although not funded, has already benefited from the organizational support of LSST DESC. 

This type of large collaboration does require stable funding support. However, funding dark matter science with Rubin LSST is really a low-cost, low-risk, but high-reward investment. Dark matter science with Rubin shares similar technical and analysis needs as dark energy science. In addition, many experimental particle physicists, observational cosmologists, theorists, and simulators are already working together on dark energy science. 
Recognizing that Rubin LSST can deliver dark matter science is the best way to ensure the existing investment toward studies of systematics, complex analysis pipelines, and technical work at the pixel and instrument level will also encompass important dark matter science, and hence maximize the scientific output.
Importantly, the co-location of research efforts in dark matter, dark energy, neutrinos, and inflation provides the greatest opportunity to leverage technical and scientific overlap and to capitalize on emerging scientific opportunities in the next decade.

\bibliographystyle{JHEP.bst}
\bibliography{main.bib}

\end{document}